\documentclass[a4paper,12pt]{article}
\usepackage[pctex32]{graphics}

\newcommand{\be}{\begin{equation}}
\newcommand{\ee}{\end{equation}}
\newcommand{\ben}{\begin{eqnarray}\displaystyle}
\newcommand{\een}{\end{eqnarray}}

\begin{titlepage}

\begin{document}
{\baselineskip20pt

%\begin{flushright}
%\today
%\end{flushright}

\vskip .6cm

\begin{center}
{\Large \bf Thermodynamics of Einstein-Born-Infeld black holes in three dimensions}

\end{center} }

\vskip .6cm
 \centerline{\large Yun Soo Myung$^{1,a}$,
 Yong-Wan Kim $^{1,b}$,
and Young-Jai Park$^{2,c}$}

\vskip .6cm

\begin{center}
{$^{1}$Institute of Basic Science and School of Computer Aided
Science,
\\Inje University, Gimhae 621-749, Korea \\}

{$^{2}$Department of Physics, Sogang University, Seoul 121-742, Korea}
\end{center}

\vspace{5mm}

%\renewcommand{\thefootnote}{\arabic{footnote}}
%\setcounter{footnote}{0} \setcounter{page}{2}
%\section{Introduction}

%\centerline{\bf Abstract} \bigskip
\begin{abstract}
We show that all thermodynamic quantities of the
Einstein-Born-Infeld black holes in three dimensions can be
obtained from  the dilaton and its potential of two dimensional
dilaton gravity through the dimensional reduction. These are all
between non-rotating uncharged BTZ black hole (NBTZ) and charged
BTZ black hole (CBTZ).

\end{abstract}


\vskip 0.8cm

\noindent $^a$ysmyung@inje.ac.kr \\
\noindent $^b$ywkim65@gmail.com \\
\noindent $^c$yjpark@sogang.ac.kr

\noindent
\end{titlepage}

\setcounter{page}{2}

\section{Introduction}

In 1934 Born and Infeld proposed a nonlinear electrodynamics
giving a finite value for the self-energy of a pointlike
charge~\cite{BI}. Although it became less popular with the
introduction of QED, it has been observed that the Born-Infeld
action in the presence of scalar fields arises as an effective
action governing the dynamics of vector fields on
D-branes~\cite{Tsey}. For various motivations, extending the
Reissner-Nordstr\"om (RN) black hole solutions in Einstein-Maxwell
theory to the charged black hole solutions in Einstein-Born-Infeld
(EBI) theory with/without a dilaton field has attracted much
attention in recent years, for example, see~\cite{BIW,Dey,She}. In
particular, Einstein gravity in (2+1)
dimensions~\cite{Car,Man,FHR} has been intensively studied because
of the existence of black holes solutions in (2+1)-anti de Sitter
(AdS) spacetimes~\cite{btz,btz1,mtz}, which possess certain
features inherent to the (3+1)-black holes. Moreover, Cataldo and
Garcia had obtained EBI solution for certain range of the
parameters mass, charge, cosmological and the Born-Infeld
constants, which represent a static circularly symmetric black
hole in (2+1) dimensions, although they did not analyze
thermodynamic quantities such as heat capacity and free
energy~\cite{CG}.

On the other hand, two dimensional (2D) dilaton gravity, which
naturally induces the dilaton potential from dimensional
reduction, has been used in various situations as an effective
description of 4D gravity after a black hole in string theory has
appeared \cite{witten,wit1}. In particular, thermodynamics of this
black hole has been analyzed by several authors
\cite{crff,crf1,crf2,gkv}. Another 2D theories, which were
originated from the Jackiw-Teitelboim (JT) theory
\cite{jackiw,Teit}, have been also studied
\cite{JT-theories,JT2,fnn}. Actually, the 2D dilaton gravity
approach is the good $s$-wave approximation to the higher
dimensional gravity~\cite{NO}. Recently, we have introduced the 2D
dilaton gravity approach, which completely preserves the
thermodynamics of 4D black hole without kinetic term~\cite{mod},
showing that the 2D curvature scalar shows the features of
extremal and Davies' points of RN black hole  clearly. Very
recently, we have  resolved the issue of Ruppeiner's geometric
approach ~\cite{Rup1} by analyzing the RN-AdS black
holes~\cite{Rmkp,Amkp}.

It is worth while to point out the literature on the (2+1)-EBI
gravity in 2D dilaton gravity approach through the dimensional
reduction is rather scarce. Also, it is still believed that
(2+1)-gravity will provide new insights towards a better
understanding of the (3+1)-dimensional classical and quantum
gravity. In particular, the structure of (2+1)-EBI gravity is much
simpler than that of (3+1)-EBI gravity. As far as we know, it
remains  a highly nontrivial task to carry out a completely
thermodynamic analysis of (3+1)-EBI black holes~\cite{BIW,Dey}. In
this paper, we address this issue for the  EBI black holes in
(2+1) dimensions. We will show that all thermodynamic quantities
of the EBI black holes can be systematically obtained from the  2D
dilaton gravity approach through the dimensional reduction. In
addition, we find a possible phase transition between NBTZ and
CBTZ.

The organization of this work is as follows. In Sec. 2, we briefly
review the procedure to obtain the EBI black hole solution.  In
Sec. 3, we sketch another approach of 2D dilation gravity to find
the EBI black hole solution. In Sec. 4, we investigate all
thermodynamic properties of the EBI black holes by comparing those
of NBTZ and CBTZ.  We summarize our results in Sec. 5. Finally, we
derive the ADM mass of EBI black hole by making use of  the
Hamiltonian analysis in Appendix.

\section{EBI black hole solutions}

Now, let us consider a (2+1)-gravity coupled with nonlinear electrodynamics starting
from the EBI action
\begin{equation}\label{action}
S=\int d^3x \sqrt{-g}\left(\frac{R-2\Lambda}{16\pi}+L(F)\right),
\end{equation}
where
\begin{equation}
L(F)=\frac{b^2}{4\pi}\left(1-\sqrt{1+\frac{2F}{b^2}}\right).
\end{equation}
Here, the constant $b$ is the Born-Infeld parameter, and $\Lambda$
is the cosmological constant. Notice that this Lagrangian reduces
to the Maxwell one in the Maxwell limit of  $b^2 \rightarrow
\infty$, $L(F)\rightarrow  - F/4\pi$ with
$F=\frac{1}{4}F_{\mu\nu}F^{\mu\nu}$. Varying this action with
respect to the metric $g_{ab}$ leads to the Einstein equation
\begin{eqnarray} \label{3DEE}
&G_{ab}& + ~\Lambda g_{ab}= ~8\pi T_{ab},\\
&T_{ab}& = L(F)g_{ab} - F_{ac}F_b^c~L(F),
\end{eqnarray}
while the variation with respect to the electromagnetic potential
$A_a$ yields the electromagnetic field equation
\begin{equation} \label{3DEM}
\nabla_a(F^{ab}L_{,F})=0.
\end{equation}
Here the Latin index $a,b,c\cdots$ denote 3D spacetimes and
$L_{,F}$ represents the derivative of $L(F)$ with respect to $F$.
In order to obtain the EBI black hole solution, we consider a
static circular metric in (2+1)dimensions as follows
\begin{equation}\label{3metric}
ds^2_{3D} =g_{ab}dx^a
dx^b=-U(r)dt^2+\frac{dr^2}{U(r)}+r^2d\theta^2,
\end{equation}
where $U(r)$ is the unknown metric function. Here we choose the
electric field  only to be
\begin{equation}
F_{ab}= E(r)(\delta^t_a \delta^r_b - \delta^r_a \delta^t_b).
\end{equation}
Since the invariant is given by $F = - E^2(r)/2$, from Eq. (\ref{3DEM})
we obtain the solution for electric field
\begin{equation} \label{3DEF}
E(r) = \frac{Q}{\sqrt{{r}^2 + Q^2/b^2}}.
\end{equation}
Here $Q$ is an integration constant to represent the charge. In
the Maxwell limit, we recover a correct form of $E_{ML}=Q/r$ in
(2+1) dimensions. Hence the electric field of BI theory $E(r)$ in
Eq. (\ref{3DEF}) represents the smeared version of $E_{ML}=Q/r$.
Plugging the electric configuration into the
($\theta,\theta$)-component of  Einstein equation (\ref{3DEE})
with $\Lambda=-1/l^2$ leads to
\begin{equation}
U_{,r}=\frac{2r}{l^2}-\frac{4Q^2}{\sqrt{r^2+Q^2/b^2}}-4b^2r\Bigg(\frac{r}{\sqrt{r^2+Q^2/b^2}}-1\Bigg).
\end{equation}
Then one finds the metric function after the integration of
$\int^r_{r_0} U(\tilde{r})_{,\tilde{r}}d\tilde{r}$
\begin{eqnarray} \label{EBIS}
U(r)&=&-M+\frac{r^2}{l^2}+2b^2
r\Bigg(r-\sqrt{r^2+\frac{Q^2}{b^2}}\Bigg)-2Q^2\ln\Bigg[r+\sqrt{r^2+\frac{Q^2}{b^2}}\Bigg]\\
 &+&2Q^2\ln\Bigg[l+\sqrt{l^2+\frac{Q^2}{b^2}}\Bigg]- 2 b^2 l\left[l-\sqrt{l^2+\frac{Q^2}{b^2}}\right]\nonumber.
\end{eqnarray}
Here we choose $r_0=l$ for our purpose. This solution is different
from the old solution (20) in Ref. \cite{CG} by two terms in the
second line. In fact, the old solution has some ambiguity.  In the
Maxwell limit of $b^2\to\infty$, it could not recover the charged
BTZ black hole solution~\cite{mtz}
\begin{equation}
U_{CBTZ}(r)=-M+\frac{r^2}{l^2}-2Q^2 \ln \Big(\frac{r}{l}\Big).
\end{equation}
Also, counting  the mass dimensions of $[Q] = 0,[b] = 1$, and $[l]
= - 1$, we find that the log-term in the first line has ``$ - 1$"
dimension, which contradicts to ``zero" dimension. This is because
they did not evaluate the lower bound correctly, like as $\int^r
U(\tilde{r})_{,\tilde{r}}d\tilde{r}$.  However, these two
disappear in our solution because we evaluate it correctly, as
$\int^r_{r_0} U(\tilde{r})_{,\tilde{r}}d\tilde{r}$ with $r_0=l$.
The last term in the second line comes from the third term of the
first line with $r=r_0$, while the first term in the second line
gives a correct zero dimension when combining it with the last
term in the first line. The relevant parameters for EBI black
holes are $Q$ and $ b$. The mass of EBI black holes is obtained
from $U(r)=0$ as the function of horizon radius $r$,
\begin{eqnarray} \label{mass}
M(r)&=&\frac{r^2}{l^2}+2b^2
r\Bigg(r-\sqrt{r^2+\frac{Q^2}{b^2}}\Bigg)-2Q^2\ln\Bigg[r+\sqrt{r^2+\frac{Q^2}{b^2}}\Bigg]\\
 &+&2Q^2\ln\Bigg[l+\sqrt{l^2+\frac{Q^2}{b^2}}\Bigg]-2 b^2 l\left[l-\sqrt{l^2+\frac{Q^2}{b^2}}\right]\nonumber.
\end{eqnarray}

\section{2D dilaton gravity approach}

Now, we are in a position to derive the EBI black hole solution $U(r)$
in Eq. (\ref{EBIS}) from the dilaton gravity approach.

Assuming ${\cal M}_3={\cal M}_2 \times S^1$ for our purpose, we
perform a Kaluza-Klein reduction
\begin{equation}\label{2metric}
ds^2_{3D} =g_{ab}dx^a dx^b=\bar{g}_{\mu\nu}d\bar{x}^\mu
d\bar{x}^\nu +\phi^2(\bar{x})d\theta^2 ,
\end{equation}
where $\phi$ represents the radius $r$ of circle $S^1$. Here the
Greek indices $\mu,\nu,\cdots$ represent the two-dimensional
spacetimes.  After the dimensional reduction by integrating Eq.
(\ref{action}) over $S^1$, the effective 2D dilaton action is
given by
\begin{equation}
S_{2D} = 2\pi \int d^2\bar{x}
\sqrt{-\bar{g}}\phi\left(\frac{\bar{R}_2-2\Lambda}{16\pi}+L(F)\right).
\end{equation}
Here, $\bar{R}_2$ is the 2D Ricci scalar. Note that since there
are no kinetic term for dilaton, it is not necessary to perform
conformal transformation to find the black hole thermodynamics.
This contrasts to the cases in higher dimensions than three
\cite{mod,Rmkp,Amkp}

For simplicity, by omitting 'bar', the effective field equations of the EBI theory
in two dimensions are given by
\begin{eqnarray}
\nabla^\nu\left(\frac{\sqrt{-g}\phi
F_{\mu\nu}}{\sqrt{1+\frac{2F}{b^2}}}\right)&=&0,\\
\nabla^2\phi + 2 \phi \Lambda &=&
 16\pi\phi\left(L(F) - \frac{\partial L(F)}{\partial
g^{\mu\nu}} g^{\mu\nu}\right),\\
R_2 - 2\Lambda &=& - 16\pi L(F).
\end{eqnarray}
We note that $R_2$ differs from the (2+1) dimensional Ricci scalar
$R$ in Eq. (\ref{action}). Even though $R$ is singular at $r=0$,
$R_2$ is regular because it is the 2D Ricci scalar. It could be de
Sitter ($R_2>0$), flat ($R_2=0$), and anti de Sitter spacetimes
($R_2<0$). Actually  $R_2$ is related to thermodynamic quantity,
heat capacity (See Eqs. (21) and (43)). Now, as a concrete solution of EBI dynamical
equations, we present a static self-consistent solution in the 2D
dilaton approach. Then  the restricted electric field becomes to
be $F_{\mu\nu}= E(r)(\delta^t_\mu \delta^r_\nu - \delta^r_\mu
\delta^t_\nu)$. Since the invariant is given by $2F = - E^2(r)$
from the electromagnetic field equations, we obtain the solution
for electric field as follows
\begin{equation}
E(\phi) = \frac{Q}{\sqrt{{\phi}^2 + Q^2/b^2}}.
\end{equation}
Considering
\begin{eqnarray}
&& L(F)=\frac{b^2}{4\pi}\left(1-\frac{b\phi}{\sqrt{b^2\phi^2+Q^2}}\right),\\
&& \frac{\partial L(F)}{\partial
g^{\mu\nu}}=-\frac{F_{\mu\rho}{F_\nu}^\rho}{8\pi\sqrt{1+2F/b^2}},
\end{eqnarray}
equations (6) and (7) take the forms
\begin{eqnarray}
&&\nabla^2\phi=V(\phi)=-2\phi\Lambda+4b^2\left(\phi-\sqrt{\phi^2+Q^2/b^2}\right), \label{newat1}\\
&&R_2=-V'(\phi)=2\Lambda+4b^2\left(\frac{\phi}{\sqrt{\phi^2+Q^2/b^2}}-1\right)\label{newat2}.
\end{eqnarray}
 These are exactly the same  equations derived from the 2D dilaton
 gravity with potential $V(\phi)$,
\begin{eqnarray}\label{repara-action}
\bar{I}_2=\int_{{\cal M}_2} dxdt \sqrt{-g}~[\phi R_2 + V(\phi)].
\end{eqnarray}

In order to solve the above two equations, we introduce the
Schwarzschild-type gauge for $g_{\mu\nu}$ as
\begin{equation} \label{gmetric}
g_{\mu\nu}={\rm diag}(-f,f^{-1}).
\end{equation}
Then, its curvature scalar takes the form
\begin{equation}
R_2 = -f'',
\end{equation}
where the prime ``$\prime$" denotes the derivative with respect to
$x$. We note  from Eq. (\ref{2metric}) that the dilaton is
independent of time $t$~($\phi=\phi(x)$).  Eqs. (\ref{newat1}) and
(\ref{newat2}) reduce to
\begin{equation} \label{eqp}
f\phi''+f'\phi'=V(\phi),~f''=\frac{dV(\phi)}{d\phi}.
\end{equation}
In addition, we have the kinetic term for $\phi$
\begin{equation} \label{pkin}
\label{phisq} ({\nabla}\phi)^2=f(\phi')^2.
\end{equation}
If one chooses the linear dilaton as the solution
\begin{equation}
\phi=x,
\end{equation}
then Eq. (\ref{eqp}) leads to
\begin{equation}
f'=V(\phi),~f''=V'(\phi),
\end{equation}
which imply that the latter is just a redundant relation. Then, we
obtain the solution to Eqs. (\ref{newat1}) and (\ref{newat2}) as
\begin{equation} \label{gle}
 ds^2_{2D}=g_{\mu\nu}dx^\mu dx^\nu=-f(\phi)dt^2+\frac{1}{f(\phi)}d\phi^2.
\end{equation}
Here the metric function $f(\phi)$ is given by
\begin{equation} \label{fvalue}
 f(\phi)=J(\phi) - {\cal C},
\end{equation}
where $J(\phi)$ is the integration of $V(\phi)$ with a reference point $\phi_0 = l$ given by
\begin{equation}
J(\phi)=\int^{\phi}_{\phi_0}V(\tilde{\phi})~d\tilde{\phi}.
\end{equation}
Also ${\cal C}$ is a coordinate-invariant constant of integration,
which will be identified with the mass $M$ of the EBI black hole
by using the Hamiltonian approach in Appendix.

Taking into account $\Lambda=-1/l^2$, we obtain essential
quantities for describing thermodynamics of EBI black hole:
dilaton potential $V$, its derivative $V'$, and its integration
$J$ as
\begin{eqnarray}
\label{v1} V(\phi) &=& \frac{2\phi}{l^2}+4b^2\left(\phi-\sqrt{\phi^2+Q^2/b^2}\right),\label{V}\\
\label{v2}V'(\phi) &=& \frac{2}{l^2}+4b^2\left(1-\frac{\phi}{\sqrt{\phi^2+Q^2/b^2}}\right),\label{V'} \\
\label{v3}J(\phi) &=& \frac{\phi^2}{l^2}+2b^2
\phi\Bigg(\phi-\sqrt{\phi^2+\frac{Q^2}{b^2}}\Bigg) - 2Q^2{\rm
ln}\left[\phi+\sqrt{\phi^2+\frac{Q^2}{b^2}}\right] \label{J}\\
       &+& 2Q^2{\rm ln}\left[l+\sqrt{l^2+\frac{Q^2}{b^2}}\right] - 2 b^2 l\left[l-\sqrt{l^2+\frac{Q^2}{b^2}}\right]. \nonumber
\end{eqnarray}
Here we confirm the relation of metric function between $U(r)$ and
$f(\phi)$ for setting $\phi=r$ as follows
\begin{equation}
U(r) = f(r),
\end{equation}
which  means  that a 2D dilaton gravity approach gives us  the EBI
black hole solution too.  We note that the mass of EBI black hole
is given by $M(r)=J(\phi)={\cal C}$ when considering
Eqs.(\ref{mass}) and (\ref{v3}). Hereafter we call this black hole
as the Born-Infeld-Banados-Teitelboim-Zanelli black hole (BIBTZ).
 Here
we mention the two limiting cases: uncharged limit of $Q=0$ and
Maxwell limit of $b\to\infty$ with $Q\neq0$.  The former provides
the non-rotating BTZ black hole (NBTZ) described by~\cite{HP}
\begin{equation}
V_{NBTZ}=\frac{2\phi}{l^2},~V'_{NBTZ}=\frac{2}{l^2},
J_{NBTZ}=\frac{\phi^2}{l^2}
\end{equation}
 and  the latter gives the
charged BTZ black hole (CBTZ) described by
\begin{equation}
V_{CBTZ}=\frac{2\phi}{l^2}-\frac{2Q^2}{\phi},~V'_{CBTZ}=\frac{2}{l^2}+\frac{2Q^2}{\phi^2},
J_{CBTZ}=\frac{\phi^2}{l^2}-2Q^2{\rm ln}\Big(\frac{\phi}{l}\Big).
\end{equation}
Their metric functions are given by
\begin{eqnarray}
f_{NBTZ}(\phi) &=&-M+J_{NBTZ}= -M + \frac{\phi^2}{l^2}, \\
f_{CBTZ}(\phi) &=& -M +J_{CTBZ}=-M + \frac{\phi^2}{l^2}-2Q^2{\rm
ln}\Big(\frac{\phi}{l}\Big).
\end{eqnarray}
This is why we use ``BTZ" in BIBTZ. We note that these two limits
exist for the whole region of $\phi$.

\section{Thermodynamics of  EBI black holes}

In order to describe thermodynamics of BIBTZ in terms of 2D
dilaton quantities of Eqs.(\ref{V})-(\ref{J}), we have to know the
extremal BIBTZ.  From Eq.(\ref{eqp}) with the extremal condition
$f'(x)=0$ (equivalently, $V(\phi)=0$), we have the extremal value
\begin{equation}
\phi_{e}=\frac{2b l^2 Q}{\sqrt{1+4b^2l^2}}.
\end{equation}
Obviously, we have two limiting cases: $\phi_e=0$ for NBTZ and
$\phi_e=Ql$ for CBTZ.  Inserting this into $J(\phi)$, we have the
extremal  mass $M_{e}$
\begin{eqnarray}
M_{e} &=& -2Q^2{\rm \ln}\left[\frac{Q}{b} \sqrt{1+4b^2l^2}\right]+
2Q^2{\rm ln}\left[l+\sqrt{l^2+Q^2/b^2}\right] \\ \nonumber
      &{}& - 2 b^2 l\left[l-\sqrt{l^2+Q^2/b^2}\right].
\end{eqnarray}
For $M > M_{e}$, there exist two horizons $\phi=\phi_{\pm}$, while
for $M = M_{e}$, we obtain the degenerate horizon $\phi=\phi_{e}$.
For thermodynamic analysis, we need the outer horizon of $\phi_+$
and hereafter we omit ``+".  On the other hand, for $M < M_{e}$,
there exists no horizon, or naked singularity (NS). In Fig. 1, we
depict the behaviors of $M$ versus $Q^2$ to understand the BIBTZ.
As is well known, the extremal point of NBTZ is just zero
(massless BTZ black hole) and the whole region belongs to NS (Fig.
1a), while in the Maxwell limit (CBTZ: Fig. 1d of $b=100$), we
find the minimal NS region~\cite{mtz}. On the other hand, there is
the NS region of BIBTZ between two limiting cases. We note that if
the electric charge $Q$ is large, the BIBTZ exits even for
negative masses. This is in sharp contrast with what happens in
four dimensions. In this sense, the BIBTZ including CBTZ is
problematic. Also we mention that the global structure of BIBTZ,
Penrose diagram is similar to that of CBTZ because BIBTZ has two
horizons in asymptotically AdS spacetimes.

\begin{figure}[t!]
   \centering
   \includegraphics{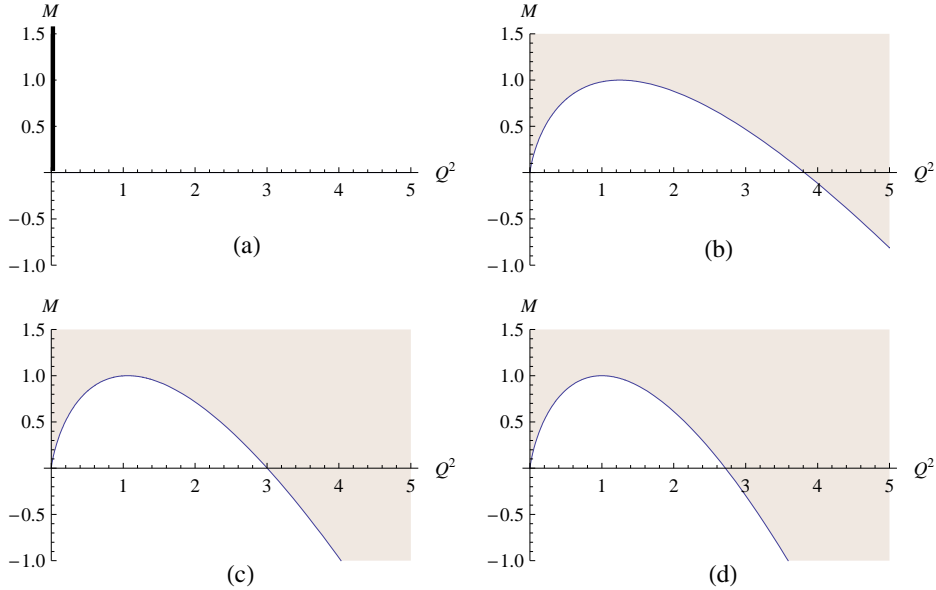}
\caption{Regions of  mass-charge plane with $l=10$. (a) $Q=0$
(NBTZ) (b) $b=0.1, Q=1$ (c) $b=0.2, Q=1$ (d) $b=100, Q=1$ (CBTZ).
Black holes exist on the positive $M$-axis for NBTZ and the shaded
areas for all BIBTZ including CBTZ. Extremal black holes are (0,0)
point for NBTZ, while these are boundary curves for BIBTZ ($b=0.1,
0.2, 100$). CBTZ has the minimal NS region (white area). }
\label{fig.1}
\end{figure}

In order to explore thermodynamics of BIBTZ, from Eqs.
(\ref{V})-(\ref{J}), we obtain the thermodynamic functions of
Hawking temperature $T(\phi)$, heat capacity $C(\phi)$, and free
energy $F(\phi)$ as follows:
\begin{eqnarray}
T(\phi)&=&\frac{V(\phi)}{4\pi}, \\
C(\phi)&=& 4\pi \Big(\frac{V(\phi)}{V'(\phi)}\Big), \\
F(\phi)&=& J(\phi) -J(\phi_e) - \phi V(\phi).
\end{eqnarray}
Note that defining the free energy, we use the extremal case as
the ground state because we are working with the canonical
ensemble of fixed-charge $Q$~\cite{CEJM}.

\begin{figure}[t!]
   \centering
   \includegraphics{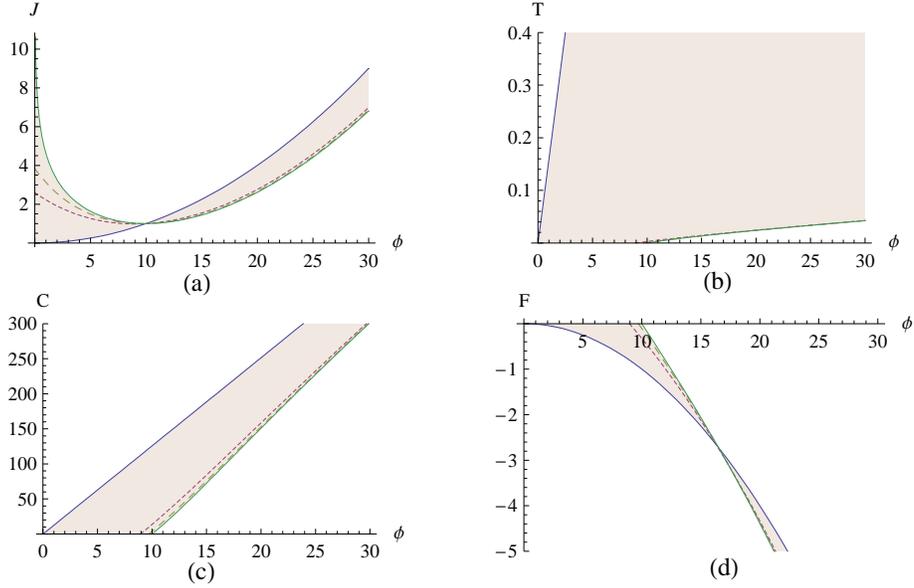}
\caption{Thermodynamic quantities for BIBTZ (shaded areas)  with
$Q=1, l=10$. (a) Mass graph $J(\phi)$.  (b) Temperature graph
$T(\phi)$. (c) Heat capacity graph $C(\phi)$. (d) Free energy
$F(\phi)$. All inner curves denote $b=0.1$ next-nearest and
$b=0.2$ nearest to $b=100$ (CBTZ).} \label{fig.2}
\end{figure}

In the Maxwell limit (CBTZ)~\cite{mtz}, we have the thermodynamic
functions as follows
\begin{eqnarray}
T_{CBTZ}(\phi)&=& \frac{1}{2\pi}\left( \frac{\phi}{l^2} - \frac{Q^2}{\phi} \right), \\
C_{CBTZ}(\phi)&=& 4\pi \left( \frac{\frac{\phi}{l^2}- \frac{Q^2}{\phi}}{\frac{1}{l^2} + \frac{Q^2}{\phi^2}}\right), \\
F_{CBTZ}(\phi)&=& - \frac{\phi^2}{{l}^2} + Q^2 - 2 Q^2 \ln
\Big(\frac{\phi}{Q l}\Big).
\end{eqnarray}
On the other hand, in the uncharged limit (NBTZ), the
corresponding thermodynamic functions are given by
\begin{equation}
T_{NBTZ}(\phi)= \frac{1}{2\pi}\frac{\phi}{l^2},~~C_{NBTZ}(\phi)=
4\pi\phi,~~F_{NBTZ}(\phi)= -\frac{\phi^2}{l^2}.
\end{equation}
Fig. 2a shows mass graph $J(\phi)$ for BIBTZ. The shaded region is
for BIBTZ. The upper boundary corresponds to CBTZ $(b=100)$ in the
region of $0\le \phi \le Ql$ and NBTZ$(Q=0)$ in the region of $
\phi \ge Ql$. The lower boundary corresponds to NBTZ in the region
of $ 0 \le \phi \le Ql$ and CBTZ in the region of $ \phi \ge Ql$.
The two masses meet at $\phi=\phi_e$ for CBTZ. This picture
implies clearly that  BIBTZ has the extremal black hole as the
minimum point of $dJ/d\phi=0$.

Temperature graph $T(\phi)$ is depicted in Fig. 2b. The shaded
region is for BIBTZ. The left boundary corresponds to NBTZ and the
bottom line of $0\le \phi \le Ql$ is for the extremal BIBTZ. The
sloping line on the bottom means the temperature of CBTZ.
Temperature of BIBTZ has the similar behavior: $T=0$ as the
extremal point and a monotonically increasing function.

Fig. 2c indicates  heat capacity graph $C(\phi)$ for BIBTZ. This
behavior is the same as that of the entropy $S(\phi)$. The shaded region is
for BIBTZ. The left boundary corresponds to NBTZ and the bottom
line is for the extremal BIBTZ. The right boundary denotes the
heat capacity of CBTZ. This means that all BIBTZ are
thermodynamically stable because of their positive heat capacity.

Finally,  free energy $F(\phi)$  is depicted in Fig. 3d. The
shaded region is for BIBTZ. This graph shows an interesting point
at $\phi=\phi_t=le^{1/2}$ which is found from the condition of
$F_{NBTZ}=F_{CBTZ}$. The left boundary corresponds to NBTZ in the
region of $0\le \phi \le \phi_t$ and CBTZ in the region of $ \phi
\ge \phi_t$. The right boundary corresponds to CBTZ in the region
of $ Ql \le \phi \le \phi_t$ and NBTZ in the region of $ \phi \ge
\phi_t$.  This may imply a second-order phase transition between
NBTZ and CBTZ~\cite{mtzs,MyungS}.

\section{Summary}

We have systematically obtained all thermodynamic quantities of
the Einstein-Born-Infeld black holes in three dimensions from the
dilaton and its potential of 2D dilaton gravity. This was
performed through the dimensional reduction by comparing those of
the Maxwell and BTZ black holes. These are all between non-rotating
uncharged black hole (NBTZ) and charged black hole (CBTZ). Actually, the
nonlinearity of BI action weakens the divergence of the curvature
scalar of CBTZ~\cite{Ida} and could connect the regular NBTZ to
singular CBTZ. Hence we clearly understand  the role of BI action
in constructing the 3D EBI black holes (BIBTZ) in the sense that
the BIBTZ plays the role of a toy model for higher dimensional EBI
black holes.

In addition, we have observed from free energy  that the
second-order phase transition may occur between NBTZ and
CBTZ~\cite{mtzs,MyungS}. For $Ql<\phi<\phi_t$, the NBTZ
configuration is more favorable than the CBTZ, while for
$\phi>\phi_t$, the CBTZ configuration is more favorable than the
NBTZ. This means that for $Ql<\phi<\phi_t$, the ground state is
the NBTZ, whereas for $\phi>\phi_t $, the ground state is chosen
to be the CBTZ.

\renewcommand{\theequation}{A\arabic{equation}}
\setcounter{equation}{0}

\section*{Appendix: ADM mass from the Hamiltonian analysis}

In this appendix, we  show explicitly that the integration
constant ${\cal C}$ in Eq. (\ref{fvalue}) is just the
Arnowitt-Deser-Missner (ADM) mass~\cite{adm} through the
Hamiltonian analysis by using the notation and convention of Refs.
\cite{Kunst1,Kunst2,Kunst3}.

Let us start with the different metric given by the
ADM-parametrization
\begin{equation} \label{hmetric}
\label{adm-metric} ds^2_{ADM}\equiv
h_{\mu\nu}dx^{\mu}dx^{\nu}=e^{2\rho}\left[-u^2dt^2+(dx+v
dt)^2\right],
\end{equation}
where $\rho$, $u$, and $v$ are functions of the coordinates
$(t,x)$. In terms of the parametrization (\ref{adm-metric}), the
action of  2D dilaton gravity in Eq.(\ref{repara-action})  is
expressed as
\begin{eqnarray}
\bar{I}_2&=&\int dt \int^{\sigma_+}_{\sigma_-} dx \left[
\frac{\dot{\phi}}{u}(2v\rho'+2v'-2\dot{\rho}) \right. \nonumber\\
&+&\left.\frac{\phi'}{u}(2u u'- 2v v'+ 2v
\dot{\rho}+2u^2\rho'-2v^2\rho') + u e^{2\rho}V(\phi)\right],
\end{eqnarray}
where the overdots and primes denote differentiation with respect
to time and space, respectively, while $\sigma_+$ ($\sigma_-$) is
the outer (inner) spatial boundary.

Following the Hamiltonian  formulation, we have the canonical
momenta
\begin{eqnarray}
\pi_\phi &=& \frac{2}{u}(-\dot{\rho}+v\rho'+v'),\\
\pi_\rho &=& \frac{2}{u}(-\dot{\phi}+v\phi'),\\
\pi_u &=& 0, \\
\pi_v &=& 0,
\end{eqnarray}
and the canonical Hamiltonian
\begin{equation}
\label{canH} H_c=\int dx \left(v {\cal F}+u{\cal G}\right),
\end{equation}
where  ${\cal F}$ and ${\cal G}$ are defined as
\begin{eqnarray}
{\cal F} &=& \rho' \pi_\rho + \phi' \pi_\phi - \pi'_\rho \approx 0, \\
{\cal G} &=&
2\phi''-2\phi'\rho'-\frac{1}{2}\pi_\rho\pi_\phi-e^{2\rho}V(\phi)
\approx 0.
\end{eqnarray}
Here, $\approx$ denotes that it is weakly vanishing on the
constraint surfaces in the sense of Dirac
formulation~\cite{dirac}. Note that $\pi_u$, $\pi_v$ (${\cal F}$,
${\cal G}$) are primary (secondary) constraints, and $u$, $v$
play the role of the Lagrange multipliers.

We  consider the linear combination of the constraints as
\begin{eqnarray}
{\cal E} \equiv -e^{-2\rho}\left(\phi'{\cal G}+
              \frac{1}{2}\pi_\rho{\cal F}\right)\approx 0.
\end{eqnarray}
Since ${\cal E}$ commutes with both the constraints ${\cal F}$ and
${\cal G}$, it is a physical quantity in the Dirac's sense.
Moreover, it can be rewritten by a  spatial derivative as
\begin{eqnarray}
  {\cal E} = \frac{\partial}{\partial x}\left[e^{-2\rho}(\frac{1}{4}\pi^2_\rho-\phi'^2)
             +J(\phi)\right] \equiv \frac{\partial {{\cal M}}}{\partial
             x}.
\end{eqnarray}
Here we have defined the terms in the square bracket as  the mass
observable ${\cal M}$.

At this stage, we show that ${\cal M}$ is the constant of  ${\cal
C}$ in Eq.(\ref{fvalue}) on the constraint surface. The constraint
surface implies that the off-diagonal form of $h_{\mu\nu}$ in
Eq.(\ref{hmetric}) reduces to the diagonal form of $g_{\mu\nu}$ in
Eq.(\ref{gle}). In this case, we have the connection
\begin{equation}
{\cal M}|_{cs}=-(\nabla \phi)^2+J(\phi)={\cal C},
\end{equation}
where we used Eqs.(\ref{pkin}) and (\ref{fvalue}).

Making use of the linear combination of the constraints, we can
further rewrite the canonical Hamiltonian as
\begin{equation}
\label{mod-ham} H_c=\int dx \left( -\tilde{u}{\cal
M}'+\tilde{v}{\cal F}\right) +H_+-H_-,
\end{equation}
where $\tilde{u}$, $\tilde{v}$ are given by
\begin{equation}
\tilde{u}=\frac{u}{\phi'} e^{2\rho},
~~\tilde{v}=v-\frac{u}{2\phi'}\pi_\rho.
\end{equation}
Here, $H_+$, $H_-$ are boundary terms which make the surface terms
in the variation of $H_c$ vanish for a given boundary condition.
As a boundary condition, if we choose the value of the dilaton
fixed at $u_+$ and make it to be time independent, the time
independence of the dilaton implies that $\tilde{v}_+=0$, or,
equivalently, $\pi_\rho|_{\sigma_+}=2v\phi'/u|_{\sigma_+}$.

As a result, we obtain the canonical Hamiltonian given by  the
first term in Eq. (\ref{mod-ham}) and $H_+$. The variation of the
resulting Hamiltonian is simply written as
\begin{equation}
\label{varH} \delta H_+({\cal M})=\tilde{u}\delta {\cal
M}|_{\sigma_+}.
\end{equation}
By solving ${\cal M}$ and $\pi_\rho|_{\sigma_+}$ at $u_+$ in terms
of $\tilde{u}$ as
\begin{equation}
\tilde{u}^2=\frac{h_{tt}}{{\cal M}-J(\phi_+)},
\end{equation}
Eq. (\ref{varH}) can be easily integrated to give
\begin{equation}
H_+({\cal M})=2\sqrt{-h_{tt}J(\phi_+)}\left(1-\sqrt{1-\frac{{\cal
M}}{J(\phi_+)}}\right).
\end{equation}
Note that the integration constant is fixed so that
$H_+\rightarrow 0$ as ${\cal M}\rightarrow 0$, and $h_{tt}$ is
also fixed at the boundary. $H_+({\cal M})$ is the dilaton gravity
analogue of the Brown-York quasi-local energy~\cite{BY}.

Finally, as $\phi_+$ goes to the infinity, one has
\begin{equation}
H_+({\cal M})\rightarrow \sqrt{\frac{-h_{tt}}{J(\phi_+)}}{\cal M}.
\end{equation}
Therefore, this gives to the ADM mass, $H_{ADM}\to {\cal M}$,
providing the physical metric is normalized to one at spatial
infinity \cite{Kunst3}. The constant of
$\sqrt{\frac{-h_{tt}}{J(\phi_+)}}$ may be related to the symmetry
of the 2D dilaton gravity.  The form of the metric (\ref{gle}) is
preserved by the following global rescalings: $\phi \to a \phi,~f
\to a^2 f, t \to t/a$. Then,  the mass seems to be $a^2 {\cal C}$,
and not ${\cal C}$. Actually, any function of ${\cal C}$ can be
called mass basing on the only relevant argument that ${\cal C}$
is coordinate invariant. In general, the mass of a black hole is
the value of the Hamiltonian calculated for the black hole
solution, and this calculation involved a careful analysis of
asymptotic behavior of the metric.  However, the ADM mass of
$H_+({\cal M})$ is not determined completely even after the
Hamiltonian calculation.

\medskip
\section*{Acknowledgments}
Y. S. Myung  was supported by the Korea Research Foundation (KRF-2006-311-C00249)
funded by the Korea Government (MOEHRD). Y.-W. Kim was supported by the Korea
Research Foundation Grant funded by Korea Government (MOEHRD):
KRF-2007-359-C00007. Y.-J. Park was supported by
the Korea Science and Engineering Foundation (KOSEF) grant
funded by the Korea government (MOST) (R01-2007-000-20062-0).

\end{document}